\documentstyle[11pt]{article}
\input{psfig}

\newcommand{\chooose}[2]{\left( \begin{array}{c} #1 \\ #2 \end{array} \right)}

\newtheorem{theorem}{\sc Theorem}
\newtheorem{lemma}{\sc Lemma}
\newtheorem{coro}{\sc Corollary}
\newtheorem{nota}{\sc Notation}
\newtheorem{defin}{\sc Definition}
\newtheorem{cla}{\sc Claim}
\newtheorem{ex}{\sc Example}

\newenvironment{proof}{\par \sc Proof.\rm}{\hspace*{\fill}$\Box$\vspace{1ex}}

\newenvironment{claim}{\begin{cla}}{\end{cla}}

\newenvironment{definition}{\begin{defin}}{\end{defin}}

\title{New Applications of the Incompressibility Method: Part I}
\author{Tao Jiang\thanks{Supported in part by the NSERC
Research Grant OGP0046613 and a CGAT grant.
Address: Department of Computer Science,
McMaster University, Hamilton, Ont L8S 4K1, Canada.
Email: jiang@maccs.mcmaster.ca}\\
McMaster University
\and
Ming Li\thanks{
Supported in part by
the NSERC Research Grant OGP0046506, CITO, a CGAT grant, and the
Steacie Fellowship. Address:
Department of Computer Science, University of Waterloo,
Waterloo, Ont. N2L 3G1, Canada. E-mail: mli@math.uwaterloo.ca}\\
University of Waterloo
\and
Paul Vit\'{a}nyi\thanks{
Partially supported by the European Union
through NeuroCOLT ESPRIT Working Group Nr. 8556,
and by  NWO through NFI Project ALADDIN
number NF 62-376.
Address: CWI,
Kruislaan 413, 1098 SJ Amsterdam, The Netherlands.
Email: paulv@cwi.nl}\\
CWI and University of Amsterdam}
\date{}
 
\begin{document}
\maketitle
 
\begin{abstract}
The incompressibility method is an elementary yet powerful 
proof technique. It has been used successfully in many
areas~\cite{LiVi93}. To further
demonstrate its power and elegance
we exhibit new simple 
proofs using the incompressibility method.
\end{abstract}
 
\section{Introduction}
The incompressibility of individual random objects 
yields a simple but powerful proof technique:
{\em the incompressibility method}.
This method is a general purpose
tool that can be used to prove lower bounds on computational
problems, to obtain combinatorial properties of concrete objects, and
to analyze the average complexity of an algorithm. Since the early
1980's, the incompressibility method has been successfully used 
to solve many well-known questions that had been open for a
long time and to supply new
simplified proofs for known results. A survey is \cite{LiVi93}.

The purpose of this paper is pragmatic, in
the same style as \cite{LiVi94}, and a companion
paper~\cite{BJLV98}. We want to 
further demonstrate how easy the incompressibility method can be used, via a 
new collection of simple examples. The proofs we have chosen to be included
here are not difficult ones. They are from diverse
topics. Most of these are well-known topics such
as sorting. Some results are new (but this
is not important) such as curve fitting lower bound, and
some results were known before.
In all cases, the new 
proofs are much simpler than the old ones (if they exist).

\section{Kolmogorov Complexity and the Incompressibility Method}
We use the following notation.
Let $x$ be
a finite binary string. Then $l(x)$ denotes the {\em length}
(number of bits) of $x$. In particular, $l(\epsilon)=0$
where $\epsilon$ denotes the {\em empty word}.

We can map $\{0,1\}^*$ one-to-one onto the
natural numbers by associating each string with its index
in the length-increasing lexicographical ordering
\begin{equation}
( \epsilon , 0),  (0,1),  (1,2), (00,3), (01,4), (10,5), (11,6),
\ldots .
\label{(2.1)}
\end{equation}
This way we have a binary representation for the
set of all natural numbers that is different from the standard
binary representation.
It is convenient not to distinguish between the
first and second element of the same pair,
and call them ``string'' or ``number''
arbitrarily. As an example, we have $l(7)=00$.
Let $x,y, \in {\cal N}$, where
${\cal N}$ denotes the natural
numbers. 
Let $T_0 ,T_1 , \ldots$ be a standard enumeration
of all Turing machines.
Let $\langle \cdot ,\cdot \rangle$ be a standard one-one mapping
from ${\cal N} \times {\cal N}$
to ${\cal N}$, for technical reasons chosen such that
$l(\langle x ,y \rangle) = l(y)+O(l(x))$.

Informally, the Kolmogorov complexity, \cite{Ko65},
of $x$ is the length of the
{\em shortest} effective description of $x$.
That is, the {\em Kolmogorov complexity} $C(x)$ of
a finite string $x$ is simply the length
of the shortest program, say in
FORTRAN (or in Turing machine codes)
encoded in binary, which prints $x$ without any input.
A similar definition holds conditionally, in the sense that
$C(x|y)$ is the length of the shortest binary program
which computes $x$ on input $y$. 
Kolmogorov complexity is absolute in the sense
of being independent of the programming language,
up to a fixed additional constant term which depends on the programming
language but not on $x$. We now fix one canonical programming
language once and for all as reference and thereby $C()$.
For the theory and applications, as well as history, see \cite{LiVi93}.
A formal definition is as follows:

\begin{definition}
\rm
Let $U$ be an appropriate universal Turing machine
such that 
\[U(\langle \langle i,p \rangle ,y \rangle ) =
T_i (\langle p,y\rangle) \]
 for all $i$ and $\langle p,y\rangle$.
The {\em conditional Kolmogorov complexity} of $x$ given $y$
is
\[C(x|y) = \min_{p \in \{0,1\}^*} \{l(p): U (\langle p,y\rangle)=x \}. \]
The unconditional Kolmogorov complexity of $x$ is defined
as $C(x) := C(x| \epsilon )$.
\end{definition}
It is easy to see that there are strings that can be described
by programs much shorter than themselves. For instance, the
function defined by $f(1) = 2$ and $f(i) = 2^{f(i-1)}$
for $i>1$ grows very fast, $f(k)$ is a ``stack'' of $k$ twos.
Yet for each $k$ it is clear that $f(k)$
has complexity at most $C(k) + O(1)$.

By a simple counting argument one can show
that whereas some strings can be enormously compressed,
the majority of strings can hardly be compressed
at all.
For each $n$ there are $2^n$ binary
strings of length $n$, but only
$\sum_{i=0}^{n-1} 2^i = 2^n -1$ possible shorter descriptions.
Therefore, there is at least one binary string
$x$ of length $n$ such that $C(x)   \geq   n$.
We call such strings $incompressible$. It also
follows that for any length $n$ and any binary string $y$,
there is a binary string $x$ of length $n$ such that
$C(x| y)   \geq   n$.
\begin{definition}
\rm
For each constant $c$ we say a string $x$ is
\it c-incompressible
\rm if $C(x)   \geq   l(x) -c$.
\label{string!incompressibility of}
\end{definition}
 
Strings that are incompressible (say, $c$-incompressible
with small $c$) are patternless,
since a pattern could be used to reduce
the description length. Intuitively, we
think of such patternless sequences as being random, and we
use ``random sequence'' synonymously with ``incompressible sequence.''
It is possible to give a rigorous formalization of the intuitive notion
of a random sequence as a sequence that passes all
effective tests for randomness, see for example \cite{LiVi93}.
 
How many strings of length $n$ are $c$-incompressible?
By the same counting argument we find that the number
of strings of length $n$ that are $c$-incompressible
is at least $2^n - 2^{n-c} +1$. Hence
there is at least one 0-incompressible string of length $n$,
at least one-half of all strings of length $n$ are 1-incompressible,
at least three-fourths  of all strings
of length $n$ are 2-incompressible, \ldots , and
at least the $(1- 1/2^c )$th part
of all $2^n$ strings of length $n$ are $c$-incompressible. This means
that for each constant $c   \geq   1$ the majority of all
strings of length $n$ (with $n   >   c$) is $c$-incompressible.
We generalize this to the following simple but extremely
useful %
{\it Incompressibility Lemma}.
\begin{lemma}
\label{C2}
Let $c$ be a positive integer.
For each fixed $y$, every
set $A$ of cardinality $m$ has at least $m(1 - 2^{-c} ) + 1$
elements $x$ with $C(x| y)   \geq   \lfloor \log m \rfloor  - c$.
\end{lemma}
\begin{proof}
By simple counting.
\end{proof}

Note that obviously, for some $c$, 
$C(x|y,A)\leq\left\lfloor\log m \right\rfloor +c$.

As an example for the lemma, 
set $A =  \{ x: l(x) = n  \}  $. Then the cardinality
of $A$ is $m = 2^n$.
Since it is easy to assert that $C(x)  \leq n + c$ for some
fixed $c$ and all $x$ in $A$, Lemma~\ref{C2} demonstrates
that this trivial estimate is quite sharp. The deeper
reason is that since there are few short programs, there can
be only few objects of low complexity.

\begin{definition}
\rm
  A {\em prefix set}, or prefix-free code, or prefix code, is a set of
strings such that no member is a prefix of any other member.
  A prefix set which is the domain of a partial recursive function
(set of halting programs for a Turing machine) is a special type of
prefix code called a {\em self-delimiting} code because there is an
effective procedure which reading left-to-right 
determines where a code word ends without
reading past the last symbol.
  A one-to-one function with a range that is a self-delimiting code
will also be called a self-delimiting code.
\end{definition}

A simple self-delimiting code we use throughout is obtained by reserving one
symbol, say 0, as a stop sign and encoding a
natural number $x$ as $1^x 0$.
We can prefix an object with its length and iterate
this idea to obtain ever shorter codes:
\begin{equation}
\label{ladder}
E_i (x)  = \left\{ \begin{array}{ll}
1^x 0 & \mbox{for $i=0$}, \\
E_{i-1} (l(x)) x & \mbox{for $i>0$}.
\end{array} \right.
\end{equation}
Thus, $E_1 (x) = 1^{l(x)} 0 x$ and has
length $l(E_1 (x)) = 2l(x) + 1$; 
$E_2 (x)  =  E_1 (l(x)) x$ and has length
$ l(E_2 (x) )  =  l(x) + 2l(l(x)) + 1$.
   We have for example
 \[
  l(E_3 (x))\leq l(x)+\log l(x) + 2\log\log l(x) + 1.
 \]

  Define the pairing function
 \begin{equation}\label{e.ngl}
  \langle x,y \rangle =E_2(x)y
 \end{equation}
  with inverses $\langle \cdot \rangle _1,\langle\cdot\rangle_2$.
This can be iterated to
$\langle  \langle \cdot , \cdot \rangle , \cdot \rangle$.

In a typical proof using the incompressibility method,
one first chooses an individually random object from the
class under discussion.
This object is effectively incompressible.
The argument invariably says that if a desired property
does not hold, then the object
can be compressed. This yields the required contradiction.
Then, since most objects are random, the desired property 
usually holds on average.

\section{Lower Bound for Sorting}

We begin this paper with a very simple incompressibility proof 
for a well-known lower bound on comparison based sorting.

\begin{theorem}
Any comparison based sorting algorithm requires
$\Omega (n \log n)$ comparisons to sort an array of $n$ elements.
\end{theorem}

\begin{proof}
Let $A$ be any comparison based sorting algorithm. Consider
permutation $I$ of $\{1, \ldots ,n \}$ such that
\[
C(I | A,P ) \geq \log n!
\]
where $P$ is a fixed program to be defined.
Suppose $A$ sorts $I$ in $m$ comparisons.
We can describe $I$ by recording the binary outcomes of the $m$
comparisons, which requires a total of $m$ bits. 
Let $P$ be such a program converting $m$ to $I$.
Thus,
\[
 m \geq C(I|A,P) \geq \log n!
\]
Hence, $m \geq \log n! = n\log n - O(n)$.  

\end{proof}

The above proof also easily implies a lower bound of $n\log n - O( n)$
on the average number of comparisons required for sorting.


\section{Space Filling Curves}

In \cite{NRS97}, Niedermeier, Reinhardt, and Sanders studied
the following problem: 
In an $n \times n$ mesh, consider a computable curve fitting scheme
that maps the numbers from $\{ 1, \ldots , n^2 \}$ into the mesh,
each number occupying one spot in the mesh.
Many algorithms in parallel computing, computational geometry,
and image processing depend on ``locality-preserving'' indexing
scheme for meshes. \cite{NRS97} has shown that for any indexing
scheme, there exist a pair $i$ and $j$ such that
\[
d(i,j) \geq \sqrt{3.5 |i-j|} - 1
\]
where $d$ is Euclidean distance. (When $d$ is other distances,
like Manhattan or $l_\infty$, Kolmogorov complexity argument
works similarly.) However, it is much more interesting
to obtain an ``average-case'' bound, both theoretically
and practically. The question for the average-case is open.
In fact, many experiments have been performed by
researchers in order to determine the average
distance~\cite{NRS97}.
We prove such a bound here with much
simpler argument using the incompressibility method.

\begin{theorem}
$d(i,j) \geq \sqrt {2.5 |i-j|}$ for many pairs
of $i,j$'s. 
\end{theorem}

\begin{proof}
Let $N=n^2$. Consider a computable curve-fitting scheme $F$.
Let's assume that $F$ puts $i$ in a corner. 
Consider $j$'s such that 

\begin{equation}\label{def}
C(j|i) \geq \log N.
\end{equation}

We know that there is a constant $c>0$ such
that for every $N$ there are $N/c$ such $j$'s (if there exists one such
$j$, then there exist $1/c$ portion of them by the argument
used in Exercise 2.2.6, p. 117 in \cite{LiVi93}). 

Also, we can argue that $|i-j| \leq N/2$ for at least half of the $j$'s.
For if this is not the case, we can change the universal TM
in the definition of Kolmogorov complexity by just making the new
universal TM printing 0 (1) whenever the old universal TM prints 1 (0).
Then for each $j$, let $j$' be the 1's complement of $j$, we have either
\[
    |i-j|  \leq  N/2,
\]
or
\[
    |i-j'| \leq N/2.
\]

For all the $j$ satisfying Inequality~\ref{def}, if not half of them
satisfy $|i-j| \leq N/2$, we can use that new universal TM such that
more than half of the $j$' satisfy $|i-j'| \leq N/2$. 
And under the new universal TM, the $j$' satisfy Inequality~\ref{def} if
$j$ does.

Now given $i$, index $j$ can be specified in
\[
  \log \pi d(i,j)^2 
\]
bits. But since $i$ is a corner point, we only need to enumerate $1/4$
of the numbers, so to specify $j$, we really need only
\[
  \log \frac{1}{4} \pi d(i,j)^2
\]
bits. Thus,
\[
  \log \frac{1}{4} \pi d(i,j)^2  \geq C(j|i) \geq \log N,
\]
hence
\[
  d(i,j) \geq \sqrt{4N/\pi} \geq \sqrt{8|i-j|/\pi } \approx \sqrt{2.5|i-j|}
\]
for $N/2c$-many $j$'s.   

\end{proof}

Note, this applies to other distances ($l_\infty$ and Manhattan)
discussed in \cite{NRS97} as well. It is clear one can obtain
a weaker average-case bound by consider any $i$ instead of 
corner point $i$.

Question: Can we improve this bound?
Can we improve $2.5$ to close to the worst-case 3.5 constant factor in
\cite{NRS97}? The upper bound is 4 given in \cite{NRS97}.

\section{Expected Length of a Longest Common Subsequence}~\label{sec.lcs}
For two sequences ({\it i.e.} strings) $s = s_1 \ldots s_m$ and
$t = t_1 \ldots t_n$, we say that
$s$ is a {\em subsequence} of $t$ if for some $i_1 < \ldots < i_m$,
$s_j = t_{i_j}$. A {\em longest common subsequence} (LCS) of
sequences $s$ and $t$ is a longest possible sequence $u$ that is a
subsequence of both $s$ and $t$. For simplicity, we will only consider
binary sequences over the alphabet $\Sigma = \{0,1\}$.

Let $n$ be an arbitrary positive integer and consider two random strings
$s$ and $t$ that are drawn independently from the uniformly distributed
space of all binary string of length $n$. We are interested in the expected
length of an LCS of $s$ and $t$. Tight bounds on the expected LCS length
for two random sequences is a well-known open question in 
statistics~\cite{paterson,sankoff}. After a series of papers, the best
result to date is that the length is between $0.762n$ and 
$0.838n$~\cite{chvatal,dancik,deken79,deken83}. The proofs are based on
intricate probablistic and counting arguments. Here we give simple
proofs of some nontrivial upper and lower bounds using the 
incompressibility method.

\begin{theorem}\label{lcs.upper}
The expected length of an LCS of two random sequences of length $n$
is at most $0.867n + o(n)$.
\end{theorem}
\begin{proof}
Let $n$ be a sufficiently large integer. Observe that the expected length
of an LCS of two random sequences of length $n$ is trivially bounded between
$n/2 $ and $n$. By the Incompressibility Lemma, out of the $2^{2n}$ pairs of
binary sequences of length $n$, at least $(n-1)2^{2n}/n$ of them are 
$\log n$-incompressible. Hence, it suffices to consider $\log n$-incompressible
sequences.

Take a $\log n$-incompressible string $x$ of length $2n$, and let
$s$ and $t$ be the first and second halves of $x$ respectively.
Suppose that string $u$ is an LCS of $s$ and $t$. In order to relate the
Kolmogorov complexity of $s$ and $t$ to the length of $u$, we re-encode
the strings $s$ and $t$ using the string $u$ as follows. (The idea
was first introduced in~\cite{JiLi95}.)

We first describe how to re-encode $s$.
Let the LCS $u = u_1 u_2 \cdots u_m$, where $m = l(u)$. We align the bits of
$u$ with the corresponding bits of $s$ greedily from left to right, and
rewrite $s$ as follows:
\[
s = \alpha_1 u_1 \alpha_2 u_2 \cdots \alpha_m u_m s'.
\]
Here $\alpha_1$ is the longest prefix of $s$ containing no $u_1$,
$\alpha_2$ is the longest substring of $s$ following the bit $u_1$
containing no $u_2$, and so on, and $s'$ is the remaining part of $s$
after the bit $u_m$. Thus $\alpha_i$ does not contain bit $u_i$, for
$i= 1, \ldots, m$. In other words, each $\alpha_i$ is a unary string
consisting of the bit complementary to $u_i$. We re-encode $s$ as string: 
\[
s(u) = 0^{l(\alpha_1)} 1 0^{l(\alpha_2)} 1 \cdots 0^{l(\alpha_m)} 1 s'.
\]
Clearly, given $u$ we can uniquely decode the encoding $s(u)$ to obtain $s$.

Similarly, the string $t$ can be rewritten as 
\[
t = \beta_1 u_1 \beta_2 u_2 \cdots \beta_m u_m t',
\]
where each $\beta_i$ is a unary string consisting of the bit complementary
to $u_i$, 
and we re-encode $t$ as string: 
\[
t(u) =  0^{l(\beta_1)} 1 0^{l(\beta_2)} 1 \cdots 0^{l(\beta_m)} 1 t'.
\]

Hence, the string $x$ can be described by the following information
in the self-delimiting form:
\begin{enumerate}
\item A description of the above discussion.
\item The LCS $u$.
\item The new encodings $s(u)$ and $t(u)$ of $s$ and $t$.
\end{enumerate}

Now we estimate the Kolmogorov complexity of the above description of $x$.
Items 1 and 2 take $m + O(1)$ bits. Since $s(u)$ contains at least $m$ $1$'s,
it is easy to see by simple counting and Stirling approximation 
(see {\it e.g.}~\cite{LiVi93}) that 
\begin{eqnarray*}
C(s(u)) & \leq & \log \sum_{i = m}^{n} \chooose{n}{i} ~+~ O(1) \\
& \leq & \log (\frac{n}{2} \chooose{n}{m}) ~+~ O(1)\\
& \leq & \log n  + \log \chooose{n}{m} ~+~ O(1)\\
& \leq & 2\log n  + n\log n - m\log m - (n-m)\log (n-m) + O(1) 
\end{eqnarray*}
The second step in the above derivation follows from the trivial fact that
$m \geq n/2$.  Similarly, we have
\begin{eqnarray*}
C(t(u)) & \leq & 2\log n  + n\log n - m\log m - (n-m)\log (n-m)  + O(1)
\end{eqnarray*}
Hence, the above description requires a total size of
\[ O(\log n) + m + 2n\log n - 2m\log m - 2(n-m)\log (n-m). \]

Let $p = n/m$. Since $C(x) \geq 2n - \log n$, we have
\begin{eqnarray*}
2n - \log n & \leq & O(\log n) + m + 2n\log n - 2m\log m - 2(n-m)\log (n-m) \\
& = & O(\log n) + pn - 2np\log p - 2n(1-p)\log (1-p) 
\end{eqnarray*}
Dividing both sides of the inequality by $n$, we obtain
\[ 2 \leq o(1) + p - 2p\log p - 2(1-p)\log (1-p)  \]
Solving the inequality numerically we get $p \leq 0.867 - o(1)$.    
\end{proof}

Next we prove that the expected length of an LCS of two random sequences
of length $n$ is at least $0.66666n - O(\sqrt{n\log n})$.
To prove the lower bound, we will
need the following greedy algorithm for computing common subsequences
(not necessarily the longest ones).

\bigskip
{\noindent \bf Algorithm} Zero-Major($s = s_1 \cdots s_n, t = t_1 \cdots t_n$)
\begin{enumerate}
\item Let $u := \epsilon$ be the empty string.
\item Let $i := 1$ and $j := 1$.
\item Repeat steps 4-6 until $i > n$ or $ j > n$ 
\item $~~~$ If $s_i = t_j$ then append bit $s_i$ to string $u$; and
$i:=i+1$, $j:=j+1$
\item $~~~$ Elseif $s_i = 0$ then  $j := j+1$.
\item $~~~$ Else $i := i+1$.
\item Return string $u$.
\end{enumerate}

\begin{theorem}\label{lcs.lower}
The expected length of an LCS of two random sequences of length $n$
is at least $0.66666n - O(\sqrt{n\log n})$.
\end{theorem}
\begin{proof}
Again, let $n$ be a sufficiently large integer, and take a 
$\log n$-incompressible string $x$ of length $2n$. Let
$s$ and $t$ be the first and second halves of $x$ respectively.
It suffices to show that the above algorithm Zero-Major produces
a common subsequence $u$ of length at least
$0.66666n - O(\sqrt{n\log n})$ for strings $s$ and $t$.

The idea is to encode $s$ and $t$ (and thus $x$) using information
from the computation of Zero-Major on strings $s$ and $t$.
We consider the comparisons made by Zero-Major in the order that they
were made, and create a pair of strings $y$ and $z$ as follows.
For each comparison $(s_i,t_j)$ of two complementary bits, we simply append
a $1$ to $y$. For each comparison $(s_i,t_j)$ of two identical bits,
append a bit $0$ to the string $y$. Furthermore, if this comparison
of identical bits is preceded by a comparison $(s_{i'},t_{j'})$ of two
complementary bits, we then append a bit $0$ to the string $z$ if $i' = i-1$
and a bit $1$ if $j' = j-1$.
When one string ($s$ or $t$) is exhausted by the comparisons,
we append the remaining part (call this $w$) of the other string to $z$.

As an example of the encoding, consider strings $s = \mbox{1001101}$
and $t = \mbox{0110100}$. Algorithm Zero-Major produces a common
subsequence $0010$. The following figure depicts the comparisons
made by Zero-Major, where a ``\verb+*+'' indicates a mismatch and a
``\verb+|+'' indicates a match.
\begin{center} 
\begin{verbatim}
          s =               10  01101
          comparisons       *|**||*|*
          t =                01101 0 0
\end{verbatim}
\end{center}
Following the above encoding scheme, we obtain
$y = \mbox{101100101}$ and $z = \mbox{01100}$.

It is easy to see that the strings $y$ and $z$ uniquely encode $s$ and $t$
and, moreover, $l(y) + l(z) = 2n$. Since
$C(yz) \geq C(x) - 2\log n \geq 2n - 3\log n - O(1)$,
and $C(z) \leq l(z) + O(1)$, we have
\[ C(y) \geq l(y) - 3\log n - O(1) \]
Similarly, we can obtain
\[ C(z) \geq l(z) - 3\log n - O(1) \]
and
\[ C(w) \geq l(w) - 3\log n - O(1) \]
where $w$ is the string appended to $z$ at the end of the above encoding.

Now let us estimate the length of the common subsequence $u$ produced by
Zero-Major on strings $s$ and $t$. Let $\#zeroes(s)$ and $\#zeroes(t)$ be the
number of $0$'s contained in $s$ and $t$ respectively. Clearly, $u$
contains $\min\{\#zeroes(s),\#zeroes(t)\}$ $0$'s.
 From~\cite{LiVi93} (page 159), since
both $s$ and $t$ are $\log n$-incompressible, we know 
\[ n/2 - O(\sqrt{n\log n}) \leq \#zeroes(s) \leq n/2 + O(\sqrt{n\log n}) \]
\[ n/2 - O(\sqrt{n\log n}) \leq \#zeroes(t) \leq n/2 + O(\sqrt{n\log n}) \]
Hence, the string $w$ has at most $O(\sqrt{n\log n})$ $0$'s. 
Combining with the fact that $C(w) \geq l(w) - 3\log n - O(1)$ and
the above mentioned result in~\cite{LiVi93}, we claim
\[ l(w) \leq O(\sqrt{n\log n}). \]
Since $l(z) - l(w) = l(u)$, we have a lower bound on $l(u)$:
\[ l(u) \geq l(z) - O(\sqrt{n\log n}). \]

On the other hand, since every bit $0$ in the string $y$ corresponds to
a unique bit in the common subsequence $u$, we have $l(u) \geq \#zeroes(y)$.
Since $C(y) \geq l(y) - 2\log n - O(1)$, 
\[ l(u) \geq \#zeroes(y) \geq l(y)/2 - O(\sqrt{n\log n}). \]
Hence,
\[ 3 l(u) \geq l(y) + l(z) - O(\sqrt{n\log n}) \geq 2n - O(\sqrt{n\log n}). \]
That is,
\[ l(u) \geq 2n/3 - O(\sqrt{n\log n}) \approx 0.66666n - O(\sqrt{n\log n}) \]
\end{proof}

Our above upper and lower bounds are not as tight as the 
ones in~\cite{chvatal,dancik,deken79,deken83}. Recently, Baeza-Yates and 
Navarro improved our analysis and obtained a slightly better upper of
$0.860$~\cite{BaNa97}.
It will be interesting to know if stronger bounds can be obtained
using the incompressibility method 
by more clever encoding schemes. 

\section{Multidimensional Random Walks}

Consider a random walk in 1 dimension with fixed probability
$p= \frac{1}{2}$ of taking a unit step left or right.
It is well-known that the maximal distance
from the start position in either direction 
in a random walk of $n$ steps is in the order of $\sqrt n$
with high probability. For example, the {\em Law of the 
Iterated Logarithm}, \cite{Kh24}, says
that the limit superior of
 this distance equals $\sqrt{\frac{1}{2} n \log \log n }$
with probability 1 for $n$ rises unboundedly. 
Nonetheless, probabilistic analyses of random walks 
as in \cite{Fe,Re90} apparently are not concerned with 
flexible tradeoffs
between probability and absolute upper or lower bounds
on the largest distance traveled from the origin in every dimension
as the theorem below. Such results however are very useful
in theory of computation.

 In a random walk in $k>1$ dimensions
where each step increases or decreases the distance from
the origin by a unit in exactly one dimension
we would like to know the probability of traveling distance
$d$ from the origin in any dimension in $n$ steps.

\begin{theorem}
Consider a random walk in $k$ dimensions where each step
is a unit step in any (but only one at a time)  single dimension in positive or negative
direction with uniform probability $1/2k$. 
Let $\delta (\cdot) $ be a monotonic nondecreasing
function and let $x$ be a random walk of
length $n$ such that $C(x|n) > n - \delta (n)$. 
If $n \gg k$ then the random walk $x$
has all of the following properties
 (which therefore hold with probability at least $1-1/2^{\delta (n)}$ 
for a random walk of length $n$):

{\rm (i)} For every dimension, the maximal distance the walk
moves away from the starting position in either direction 
during the walk is
 $O( \sqrt{ \frac{n}{k} (\delta (n) +\log \frac{n}{k})} )$; 

{\rm (ii)} For every dimension, the maximum distance the walk 
is away from the starting
position in either direction at the end of the walk is 
 $O( \sqrt{ \delta (n) \frac{n}{k}} )$; and

{\rm (iii)} For every dimension, the minimum distance the walk 
is away from the starting
position in either direction at the end of the walk is 
$ \Omega ( \sqrt{ 2^{- \delta (n)} \frac{n}{k} } )$.

{\rm (iv)} For every dimension, the minimum distance the walk 
is away from the starting
position in either direction at the end of an initial $m$-length
segment $x'$ with $x=x'z$ for some $z$,
 $C(x'|m) > m - \delta (m)$, and $m \gg k$, is
$ \Omega ( \sqrt{ 2^{- \delta (m)} \frac{m}{k} } )$.
\end{theorem}

\begin{proof}
($k =1$) For $k=1$ we can identify left with 0 and right with 1.
So we are interested in the deviation of the relative frequency 
of 1's in random walk $x$ of length $n$. 

(i) If $C(x|n) > n -\delta(n)$
then for all $n$
\begin{equation}\label{eq.dist}
|\#ones(x)  - \frac{n}{2}| \leq \sqrt{\frac{3}{2}( \delta (n)
+O(1))n/\log e}
\end{equation}
(\cite{LiVi93}). 

The righthand side of Inequality \ref{eq.dist}
is an upper bound on  the largest distance the random walk strays
from the origin in either direction
with probability $> 1-1/2^{\delta (n)}$.
This way we know the maximum distance from the origin {\em at
the end} of a high-complexity walk of length $n$. 

(ii) We
now analyze the maximum distance reached {\em during} the walk.
A prefix $y$ of length $m \leq n$ of $x$ has complexity
$C(y|m) > m - \delta (n) - (1+ \alpha ) \min \{ \log m , \log (n-m) \}
+O(1)$
where $\alpha$ is any fixed constant greater than 0.
(Otherwise we can effectively describe $x$ given $n$ by a
program $p$ of length $C(y|m) \leq m$ and a program $r$ of length
$C(x|y) \leq n-m$. To make one of them self-delimiting it suffices
to add a prefix of length $(1+ \alpha ) \min \{ \log m , \log (n-m)
\}$.)
Substituting in Inequality \ref{eq.dist} and maximizing the minimum 
involved to $n/2$ we find

\begin{equation}\label{eq.distpref}
|\#ones(y)  - \frac{m}{2}| \leq \sqrt{\frac{3}{2}( \delta (n) + (1 +
\alpha)
\log n  +O(1))m/\log e}
\end{equation}

(iii) To obtain a lower bound on the largest distance the
random walk necessarily strays from the origin in either direction
during the walk we can do no better (using incompressibility) than
determining the distance that necessarily exists in either direction
at the {\em end} of a high-complexity random walk.
Reference \cite{LiVi93} Lemma 2.6.2 on page 160 tells us 
that if $C(x|n) > n -\delta(n)$ then
\begin{equation}\label{eq.dist2}
|\#ones(x)  - \frac{n}{2}| > 2^{- \delta (n) -O(1)} \sqrt{n}.
\end{equation}
The righthand side of Inequality \ref{eq.dist2}
is a lower bound on  the largest distance the random walk strays
from the origin in either direction
with probability $> 1-1/2^{\delta (n)}$.

(iv) It is possible that a random walk $x=x'z$ has an
initial segment $x'$ with $l(x)=n$
and $l(x')=m < n$ such that $C(x|n) < n - \frac{1}{2} \log n$ 
so that the number of 0's can be equal to the number of 1's
while 
$C(x'|m) \geq m - \delta (m)$ (\cite{LiVi93}) so that the excess
of 0's over 1's (or vice versa) is at least 
$\Omega (\sqrt{2^{-\delta (m)} m })$. 
\\

($k>1$) For $k>1$ we consider random walks as
strings over the alphabet $1_0, 1_1, \ldots ,$ $k_0, k_1$
where $i_0$ is a unit step backward in the $i$-dimension
and $i_1$ is a unit step forward. We first show (see \cite{LiVi93}, p. 418)
that if the
overall string has high complexity, then also the subsequences
over the $\{i_0,i_1\}$-alphabets have high complexity,
for all $i$ ($1 \leq i \leq k$). 
\begin{claim}\label{claim.1}
\rm
Let $\epsilon > 0$.
Consider strings over an alphabet  $ \Sigma =  \{ 1,2, \ldots , k \}$.
For some $i$, denote the total number
of occurrences of $i$'s in $y \in \Sigma^n$ by $m$.
If 
there is a constant $\delta > 0$ such that
\begin{equation}\label{eq.lb1}
C(y|k,n) \geq (n - \delta n^{2 \epsilon}) \log k  \:,
\end{equation}
then 
$|m - n/k|< n^{1/2 + \epsilon}$.
\end{claim}

\begin{proof}
There are only
$D={n \choose m}  (k-1)^{n-m}$
strings $x$ of length $n$ with $m$ occurrences out of $i$. Therefore,
one
can specify $y$ by $n,a,m$ and its index $j$, with
$l(j)= \log D$ in this ensemble. An elementary estimate by
Stirling's formula yields, for some $\delta > 0$,
\[
\log {n \choose m} (k-1)^{n-m} \leq (n - \delta n^{2 \epsilon})
\log k.
\]
\end{proof}

We can replace each element of $i \in \Sigma$ by either $i_0$
or $i_1$ to obtain a string $x$ from $y$.
We know that for every $i \in \Sigma$
the subsequence $x_i$ of $i_0,i_1$'s of length $m_i$ in $x$ satisfies
\[ C(x_i|y) \leq C(x_i|m_i) \leq m_i +O(1).\]
Assume that for every $i$ ($1 \leq i \leq k$)
 string $x_i$ has randomness deficiency $\delta_i$: 
\[ C(x_i|y) \geq m_i - \delta_i . \]
 From $y,x_1, \ldots, x_k$ we can reconstruct $x$ and vice
 versa. Hence,
since $m_i$ can be retrieved from $y$ we can delimit the subprograms
of length $ C(x_i|y)$ by giving the length of $m_i - C(x_i|y)$ so that
\begin{eqnarray*}
 C(x|y) & = & C(x_1, \ldots , x_k|y) \\
& \leq & \sum_{i=1}^k 
[ C(x_i|y) + 2 \log \delta_i] \\
&\leq & n -
\sum_{i=1}^k [ \delta_i - 2 \log \delta_i  ].
\end{eqnarray*}
If 
\begin{equation}\label{eq.lb2}
 C(x|y) \geq n -  \delta (n),
\end{equation}
then
$\delta (n) \geq \sum_{i=1}^k [ \delta_i - 2 \log \delta_i]$  and therefore
\begin{equation}\label{eq.subseqv}
C(x_i|y) > m_i -  \delta (n) - 2 \log \delta (n) 
\end{equation}
for every $i$ ($1 \leq i \leq k$).
For every $x \in \Sigma^n$ we have $C(x|n,k) \leq n \log 2k + O(1)$,
$C(x|y) \leq n + O(1)$ and $C(y|k,n) \leq n \log k + O(1)$.

Choosing $x$ such that
\[ C(x|k,n) \geq n \log 2k - \delta (n) \]
we have $C(x|k,n) \geq C(x|y)+C(y|k,n)-\delta (n)-O(1)$.
We also have $C(x|k,n) \leq C(x|y) + C(y|k,n)$ plus an additive term
to encode the delimiter between the two constituents in the right-hand
side.
This additive term is logarithmic in the randomness deficiency of
one of the terms. Therefore,
both  $C(x|y) \geq n  - \delta (n) $
and $C(y|k,n) \geq n \log k - \delta (n) $ up to additive terms logarithmic in
$\delta (n)$.
Now both Inequalities \ref{eq.lb1} and \ref{eq.lb2} are satisfied
simultaneously for $\delta (n) = o(n)$
(Inequality \ref{eq.lb1} requires 
$\delta (n) \leq \delta n^{2 \epsilon}$).
Then, by Claim~\ref{claim.1} for every $i$ ($1 \leq i \leq k$) the  
subsequence $x_i$ of $x$ over $\{i_0,i_1\}$ has length $m_i =
n/k \pm n^{1/2 }$ and Inequality \ref{eq.subseqv} holds.
Therefore, the $k>1$ case reduces to the case $k=1$ for every
dimension simultaneously.\footnote{This illustrates one of the advantages
of the incompressibility argument: the single string that has high Kolmogorov
complexity posesses {\em every} property simultaneously
that can be proved. This contrasts with properties proven to hold with high
probability: there only the strings
in the intersection of the high probability subsets 
necessarily satisfies all such properties simultaneously.}

Let $\#i_j(x)$ denote the number of occurrences of $i_j$ in $x$.
By our analysis
of the 1-dimensional case this implies that the difference between
the number of occurrences of $i_0$'s and $i_1$'s during
the random walk of length $n$ is bounded above and below
as follows:
\begin{eqnarray*}
| \#i_0 (x) - \#i_1 (x)| & \leq &
\sqrt{\frac{3}{2}( \delta (n)/2 + (1 + \alpha)
\log m_i  +O(1))m_i /\log e} \\
&=& O( \sqrt{ \frac{n}{k} (\delta (n)+\log \frac{n}{k}) } )
\end{eqnarray*}
This proves Items (i) and (ii).
by Inequality \ref{eq.distpref} and assuming $n \gg k$ for the last
equality.
By Inequality \ref{eq.dist2} it follows that
\begin{eqnarray*}
| \#i_0 (x) - \#i_1 (x)| & > & 2^{- \delta(n) /2 -O(1)} \sqrt{m_i} \\
& =  &\Omega ( \sqrt{ 2^{-\delta (n)} \frac{n}{k} } )
\end{eqnarray*}
assuming $n \gg k$ in the last equality.
This proves Items (iii) and (iv).
\end{proof}

With this approach to random walks we can
by varying the complexity of the walk (which implies varying
the probability of such a walk in the sense that low
complexity has high probability and higher complexity less
probability) regulate the possible
variation in the distance covered in the walk
(high complexity walks have precisely fixed distance while
low complexity walks have more uncertainty).

\subsection{Monopolist Game}
This approach is useful to solve the  ``Monopolist Game''
defined in \cite{WY97} as a formalization of a simplified version
of a neural network updating rule due to von der Malsburg
\cite{vdM73}. This updating rule plays a key role in 
explaining orientation selectivity in the brain.
\begin{quote}
In the {\em Monopolist Game} we start with $k$ players that are
given equal amounts $I/k$ of the total initial capital of $I$ units. The game
is divided into discrete rounds. At every round 
one of the players wins and receives $k-1$
units from the other players who each lose one unit. The players
have equal probabilities $1/k$ of winning a round.
The game terminates if all but one player has lost all of the money.
The surviving player has accumulated all the money and is called a
{\em monopolist}.
\end{quote}
Denote the players by elements of $\Sigma$ as in
Claim~\ref{claim.1}. Denoting just the
winner in each round we can write the outcome of $n$ rounds
as $x \in \Sigma^n$. 
\begin{theorem}
Consider the monopolist game with $k,I, \Sigma,x,n$ as defined above
and assume $n \gg k$.

(i) For an outcome $x$ with $C(x|k,n) \geq n \log k - \delta (k,n)$
with $\delta (k,n)= \delta n^{2 \epsilon}\log k$ ($\delta > 0$ a constant),
and hence with probability at least $1-1/2^{\delta (k,n)}$,
there is no monopolist for $n<I^{2/(1+2 \epsilon)}$.

(ii) For an outcome $x$ with 
$C(x|n,k) \geq n \log k - \epsilon \log n$ ($0 < \epsilon  < \frac{1}{4}$)
and hence with probability at least $1 - 1/n^{\epsilon}$
there is certainly a monopolist for some $n$ satisfying
$n  \leq (I/(k-2))^{2/(1-2 \epsilon)}$.
\end{theorem}

\begin{proof}
Let $n_i$ be the number of occurrences of
$i$ in $x$ where $l(x)=n$. This way $I/k + (k-1)n_i - (n-n_i)$ represents the
capital of $i$ at the end of $x$ (positive or negative). 

(i) We
are interested in the situation where there exist $n' \leq n$
such that $I/k + (k-1)n'_i - (n'-n'_i) \leq 0$ 
that is 
\begin{equation}\label{eq.mon1}
n'_i - \frac{n'}{k} \leq I
\end{equation}
for
$k-1$ elements $i \in \Sigma$. With  $C(x|k,n) > n- \delta (k,n)$ 
and $n<I^{2/(1+2 \epsilon)}$
Inequality \ref{eq.mon1} is violated for every $i \in \Sigma$ by
Claim~\ref{claim.1}.
\\

(ii) The number of strings in $\Sigma^n$ with frequencies $n_i = n/k$ of symbol $i$
($1 \leq i \leq k$) is
the
multinomial coefficient satisfying
\begin{eqnarray*}
\log {n \choose {n_1,    \ldots , n_k}} &= &
\log {{n!} \over {n_1 !  \cdots  n_k !} } \\
& \sim &  n \log k - \frac{1}{2}(k-1) \log n + \frac{k}{2} \log k + O(1) 
\end{eqnarray*}
using Stirling's approximation. This is the largest multinomial
coefficient. Therefore, the logarithm of the total 
number of strings with frequencies
$|n_i -n/k| < f(n)$ is asymptotically upper bounded by 
\begin{equation}\label{eq.multin}
g(n,k) = n \log k - \frac{1}{2}(k-1) \log n + \frac{k}{2} \log k 
+(k-1) \log f(n) + k+ O(1)
\end{equation}
and we can describe every string of this ensemble by giving its
index in at most that many bits. Consequently, if for some
$n$-length outcome $x$ of the monopolist game the complexity
satisfies $C(x|k,n) \geq g(n,k)$ then $|n_i -n/k| < f(n)$.
In particular, for $n \gg k$ and $C(x|n,k) \geq n \log k - \epsilon \log n$
we have $f(n) \leq n^{(1/2)- \epsilon}$.  

If $(\frac{n}{k} - f(n))(k-1) - (n-\frac{n}{k}-f(n)) \geq I$
then there is a monopolist by the $n$th round with certainty. This is the case
if $f(n) \leq I/(k-2)$ which is the case if $n= (I/(k-2))^{2/(1-2 \epsilon)}$.

\end{proof}

\section{Acknowledgements}
We thank Ian Munro for discussions on related subjects, 
Rolf Niedermeier for discussions on mesh indexing and their
paper \cite{NRS97}, 
Osamu Watanabe for drawing our attention to the random walk
problem, and a referee who had carefully read 
an earlier version of the paper and corrected several errors.

\end{document}